\documentclass[10pt,conference,letterpaper]{IEEEtran}

\ifCLASSINFOpdf
\else
   \usepackage[dvips]{graphicx}
\fi
\usepackage{url}
\usepackage[export]{adjustbox}

\usepackage{graphicx}
\usepackage{caption} 
\usepackage{subcaption}
\usepackage{algorithm,algpseudocode}

\algnewcommand{\Initialize}[1]{%
\Statex \textbf{Initialize:}
\Statex \hspace*{\algorithmicindent}\parbox[t]{.8\linewidth}{\raggedright #1}
}

\algnewcommand{\Forward}[1]{%
\Statex \textbf{Forward pass:}
\Statex \hspace*{\algorithmicindent}\parbox[t]{.8\linewidth}{\raggedright #1}
}
\usepackage{amsmath,amssymb}
\usepackage{hyperref}
\usepackage[keeplastbox]{flushend}
\usepackage{balance}
\usepackage{csquotes}
\usepackage[inline]{enumitem}
\usepackage[all=normal,paragraphs=tight,floats=tight,mathspacing=normal,wordspacing=tight,charwidths=tight,mathdisplays=normal,leading=tight,margins=normal]{savetrees}

\usepackage{xcolor}
\newcommand{\comment}[1]{}

\graphicspath{{figs/}}

\begin{document}
\bstctlcite{IEEEexample:BSTcontrol}


\title{Leveraging triplet loss and nonlinear dimensionality reduction for on-the-fly channel charting}

\author{
	\IEEEauthorblockN{
		Taha Yassine\IEEEauthorrefmark{1}\IEEEauthorrefmark{2}, Luc Le Magoarou\IEEEauthorrefmark{1}, Stéphane Paquelet\IEEEauthorrefmark{1} and Matthieu Crussière\IEEEauthorrefmark{1}\IEEEauthorrefmark{2}
		}
	\IEEEauthorblockA{
		\IEEEauthorrefmark{1}b$<>$com, Rennes, France
		}
	\IEEEauthorblockA{
		\IEEEauthorrefmark{2}Univ Rennes, INSA Rennes, CNRS, IETR - UMR 6164, F-35000 Rennes
	}
	}
\maketitle

\begin{abstract}
	Channel charting is an unsupervised learning method that aims at mapping wireless channels to a so-called chart, preserving as much as possible spatial neighborhoods. In this paper, a model-based deep learning approach to this problem is proposed. It builds on a physically motivated distance measure to structure and initialize a neural network that is subsequently trained using a triplet loss function. The proposed structure exhibits a low number of parameters and clever initialization leads to fast training. These two features make the proposed approach amenable to on-the-fly channel charting. The method is empirically assessed on realistic synthetic channels, yielding encouraging results.
\end{abstract}


\IEEEpeerreviewmaketitle

\section{Introduction}
\label{sec:introduction}
Signal processing has greatly benefited from the recent advent of machine learning and the growth of computing power. Deep learning in particular has been widely used in a plethora of domains and only a few are left untouched. In wireless communication, neural networks have been exploited to solve many classical problems ranging from physical layer tasks like channel estimation \cite{Balevi2020,Yassine2022} \comment{ (need citation)} and symbol decoding \cite{Samuel2017}\comment{ (a better example? need citation)} to higher level tasks like\comment{ (need other examples)} end-to-end learning of the transmit and receive chains \cite{Oshea2017}.
Another prominent example of deep learning application in wireless systems is user localization where the objective is to determine the physical locations of user equipments (UEs) solely based on their uplink channel measurements by a multi-antenna base station (BS) \cite{Wen2019}. Standard deep models require a dataset of channel measurements and their corresponding transmit locations to perform \emph{supervised} learning.

Recently, channel charting has been proposed as an \emph{unsupervised} alternative to user positioning where a charting function (or encoder) is learned in order to project high-dimensional channel vectors into a low dimensional space (a chart) that resembles the geometrical space of UEs' locations. The challenge consists in providing a map without a priori location knowledge, only using channel measurements.

\noindent {\bf Related work.} Channel charting was first introduced in \cite{Studer2018}. The approach therein builds on projecting precomputed channel features into a low-dimensional space using manifold learning techniques such as Sammon's mapping and autoencoders. Thereafter\comment{ (Subsequently)}, a number of channel charting approaches building on the same ideas\comment{ (principles)} have been proposed. In \cite{Huang2019}, autoencoders are used together with representation constraints as a way of injecting prior knowledge to further guide the learning process. In \cite{Lei2019}, the Sammon's mapping-based approach is reframed as a neural network model in a siamese configuration. In \cite{Ferrand2021}, channel charting is performed using triplet networks. In order to train the model, a database of channel vector triplets is constructed based on travel information of UEs. Similarly, \cite{Rappaport2021} explores the use of triplet networks for channel charting where an improved triplet loss is introduced in order to further distill prior information. Finally, \cite{LeMagoarou2021} proposes to use a distance measure that better preserves relative channel distances along with another manifold learning algorithm, namely Isomap \cite{Tenenbaum2000}, in order to map the channel observations to their low-dimensional representations.
It is also worth mentioning the work in \cite{LeMagoarou2021a} where a neural network with a particular structure tackles the task of user positioning. The model is implicitly based on the adapted distance measure of \cite{LeMagoarou2021} and benefits from a smart initialization. However, the training in \cite{LeMagoarou2021a} is done in a supervised fashion.

\noindent {\bf Contributions.} Building on earlier work \cite{Ferrand2021,LeMagoarou2021,LeMagoarou2021a}, a hybrid approach to channel charting is explored. A model-based neural network \cite{Shlezinger2020} with few parameters is structured as in \cite{LeMagoarou2021a}, allowing a smart initialization of its weights based on the adapted distance measure of \cite{LeMagoarou2021} jointly exploited with Isomap. The network is then trained using a triplet loss as in \cite{Ferrand2021}. Performance is evaluated on a dataset of synthetic realistic channels using classical manifold learning metrics.

The idea behind the smart initialization is to exploit as much prior information as possible in order to put the model in the best configuration prior to any optimization. This initialization naturally arises from the structure imposed on the network. Training is expected to gradually lead the network to configurations yielding even better results. To do so, the triplet loss relies on the way the channels are collected, bringing another source of information into the mix. The expected result is a reduction in the computational cost of both training and inference when compared to existing models, in an attempt to compute channel charts on the fly.

\section{Problem formulation}
\label{sec:problem_formulation}


Channel charting fall within the realm of machine learning in general, and dimensionality reduction in particular. It aims at projecting high-dimensional channel observations into a low-dimensional space, typically of 2 or 3 dimensions, in order to learn a \emph{channel chart}. In fact, physical channel models indicate that channel observations are subject to the manifold hypothesis \cite{Peyre2009}, meaning that although their original space is of high dimension, they are in reality governed by a small set of parameters. Those parameters are directly related to the spatial locations where the corresponding signals originate from \cite{LeMagoarou2021a}. In this sense, a successfully learned chart is a map between channel measurements and low-dimensional representations that preserves the local geometry of the original transmit locations.

\comment{(The following definition needs revision and is not quite accurate) }We consider a BS equipped with an antenna array of $N_r$ antennas, and one or more single antenna UEs placed on a map and sending signals to the BS on $S$ evenly spaced subcarriers and a central frequency $f_c$. We denote $\mathbf{h}_i\in \mathbb{C}^{M}$ the $i$th transmitted uplink channel vector, with $M=N_r\times S$, and $\mathbf{p}_i\in \mathbb{R}^D$ the corresponding transmit location, where $D$ corresponds to the number of retained spatial dimension (2 or 3 typically).
A database $\{\mathbf{h}_i\}_{i=1}^{N}$ of $N$ channel vectors is constructed through the collection of channel observations from different locations. Channel charting aims at learning a function $\mathcal{F}$ that takes as input the channel vectors $\mathbf{h}_i$ and produces low-dimensional representations denoted $\mathbf{z}_i$.
\begin{align*}
	\mathcal{F} \colon \mathbb{C}^M &\to \mathbb{R}^D \\
	\mathbf{h}_i &\mapsto \mathcal{F}(\mathbf{h}_i) = \mathbf{z}_i.
\end{align*}	
Note that the learned representations $\{\mathbf{z}_i\}_{i=1}^N$ are not necessarily of same dimension as the UE locations, but are typically chosen to be two-dimensional.

\section{Hybrid triplet network}
In order to conceive a model capable of learning the function $\mathcal{F}$, we rely on deep learning methods for what they offer in terms of capacity and fast adaptation to data. In particular, we adopt the philosophy of model-based deep learning \cite{Shlezinger2020} where neural network structures are guided by physical principles yielding hybrid models more capable than generic \emph{multilayer perceptrons} (MLPs) for a given task. The same philosophy has already been applied to channel estimation in an earlier work \cite{Yassine2022}.

\subsection{Distance measure}
The first step in our approach is to compute the distance matrix of the collected channel vectors. The adopted distance measure needs to correctly reflect the local spatial neighborhoods of channel observations. The traditional Euclidean distance is inadequate to the task because of its unusual behavior in high-dimensional spaces \cite{Aggarwal2001}.
A more adequate distance measure is the one introduced in \cite{LeMagoarou2021}. It is defined as follows:
\begin{equation}
	d^\star(\mathbf{h}_i,\mathbf{h}_j)^2 = 2-2\frac{\vert\mathbf{h}_i^H\mathbf{h}_j\vert}{\Vert{\mathbf{h}_i}\Vert_2\Vert\mathbf{h}_j\Vert_2}.
	\label{eq:distance}
\end{equation}
Based on the plane wave physical model, this distance measure removes the sensitivity of channel vectors to the fast variations of the global phase, making it reliable and suitable for retrieval of local neighborhoods. In addition, a normalization factor is introduced to mitigate the detrimental\comment{ (harmful?)} effect of signal intensity. This distance proved useful for channel charting \cite{LeMagoarou2021} and user positioning \cite{LeMagoarou2021a}.

\subsection{Hybrid encoder}
\label{sec:hybrid_encoder}
Now that a distance measure has been defined, one could use a generic manifold learning algorithm to project the channel vectors into the low-dimensional space. For instance, \cite{LeMagoarou2021} relies on Isomap to obtain the channel chart.
However, this approach suffers from two main issues:
\begin{enumerate}[label=(\arabic*)]
	\item Manifold learning algorithms like Isomap are incapable of directly performing out-of-sample projections, meaning that projecting new unseen samples would require recomputing the projection for the whole dataset which is very inefficient and cumbersome.
	\item These methods depend on the accuracy of the distance measure and are prone to errors due to its approximative nature.
\end{enumerate}
  
To fix the first issue, we propose the following strategy (summarized in algorithm \ref{alg:forward_pass}) that resembles the one introduced in \cite{LeMagoarou2021a}. We begin\comment{ (start?)} by selecting a small subset of $N_\text{init}$ collected channel observations that we organize in a matrix $\mathbf{D}\in \mathbb{C}^{M\times N_\text{init}}$ as column vectors. The more representative it is of the data distribution the better. We then compute its distance matrix using \eqref{eq:distance} and feed it to Isomap to produce an initial channel chart $\mathbf{Z}$ according to the method proposed in \cite{LeMagoarou2021}. For a given new observation $\mathbf{h}$ that we would like to project, we compute the modulus of its correlation\comment{ (incorrect term?)} to each one of the channel vectors in $\mathbf{D}$. We keep the $k$ largest elements of the obtained vector using the hard thresholding nonlinear operator denoted $\text{HT}_k(.)$. We normalize the result using an $l_1$-norm so that the vector elements sum up to 1. Finally, we multiply the normalized vector by the matrix $\mathbf{Z}$ which amounts to performing a weighted average of the projections of the $k$ most correlated channel vectors to the input vector $\mathbf{h}$.

\begin{algorithm}
	\caption{Proposed encoder\comment{ (improve structure)}}
	\label{alg:forward_pass}
	\begin{algorithmic}[1]
		\Require Subset of channel vectors $\mathbf{D}$, corresponding initial channel chart $\mathbf{Z}$, channel $\mathbf{h}$ to locate on the chart
	\State Correlation: $\mathbf{a} \gets \mathbf{D}^H\mathbf{h}$
	\State Modulus: $\mathbf{b} \gets \vert \mathbf{a} \vert$
	\State Hard thresholding: $\mathbf{c} \gets \text{HT}_k(\mathbf{b})$
	\State Normalization: $\mathbf{d} \gets \mathbf{c}/\Vert \mathbf{c}\Vert_1$
	\State Weighted sum: $\mathbf{z} \gets \mathbf{Z}\mathbf{d}$
	
	\Ensure $\mathbf{z}=\mathcal{F}(\mathbf{h})$
	\end{algorithmic}
\end{algorithm}

\begin{figure}[htp]
	\includegraphics[width=\columnwidth]{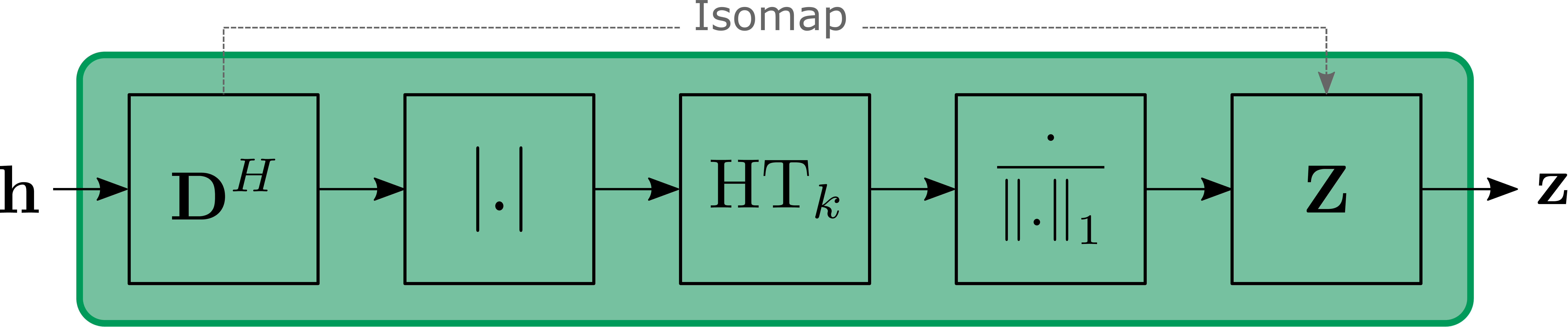}
	\caption{Hybrid encoder.}
	\centering
	\label{fig:hybrid_network}
\end{figure}

This strategy makes it possible to project individual channel observations independently of the rest of the dataset. Besides being initialized from a small dataset in conjunction with Isomap, it comprises two matrix multiplications separated by nonlinear operations which makes it easily laid out as a neural network as depicted in Fig. \ref{fig:hybrid_network}. 

The second issue is mitigated with the help of training and is tackled in the next section. Indeed, the weight matrices $\mathbf{D}$ and $\mathbf{Z}$ will be allowed to vary during the training in order to optimize the whole model.

\subsection{Triplet loss}
\label{sec:triplet_loss}
\begin{figure}[htp]
	\includegraphics[width=\columnwidth]{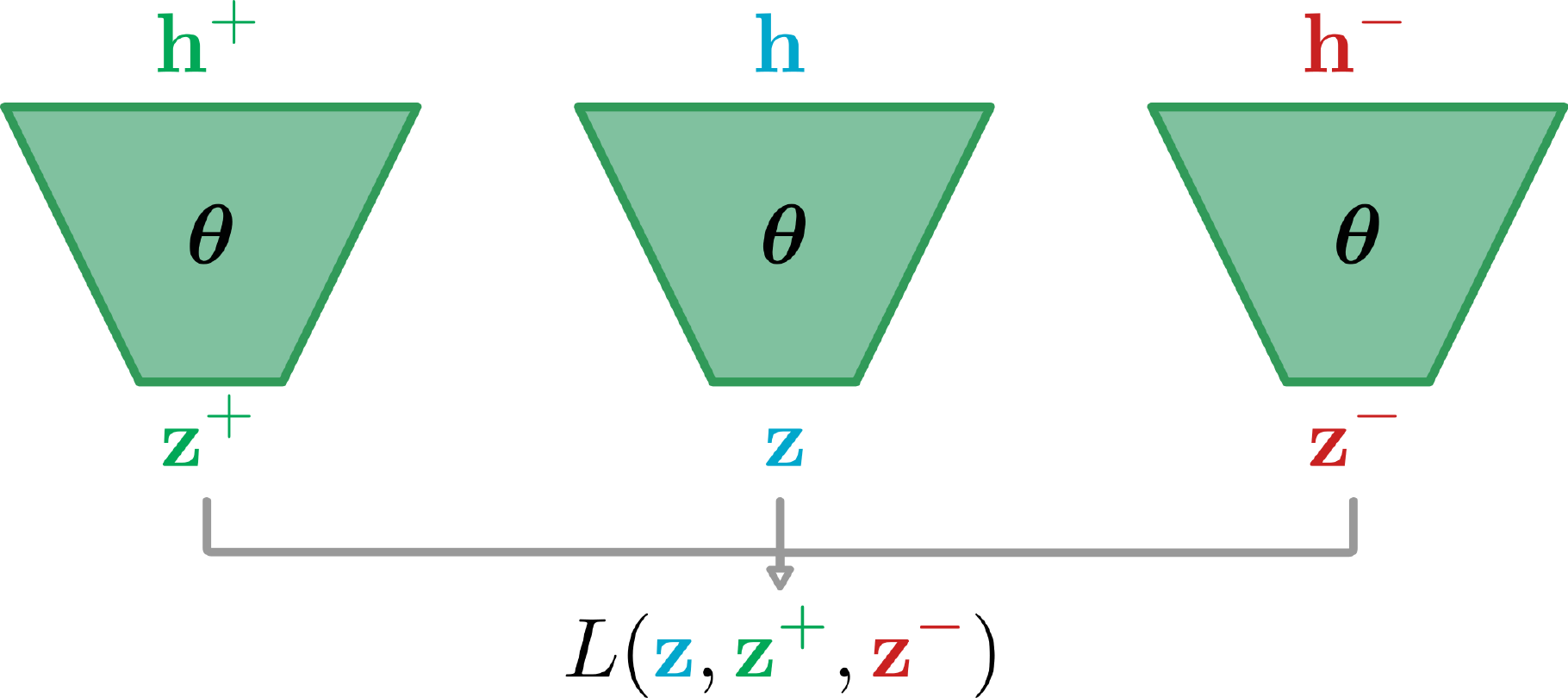}
	\caption{Triplet network structure. $\boldsymbol{\theta}$ is the set of parameters (i.e. weights) that the neural network learns throughout training.}
	\centering
	\label{fig:triplet_network}
\end{figure}

The main advantage of using neural networks is the ability to train and improve their performance through gradient descent with backpropagation \cite[Chapter~6]{Goodfellow2016}. The main idea is to calculate a loss function that quantifies the error of the network's output and to iteratively \enquote{move} its weights in the opposite direction of the gradient of this loss function as to minimize it. In a supervised learning setting, this loss function takes as input both the output of the model and the target value that is ought to be achieved, hence the need of a training dataset of target values. However, in the context of channel charting, the objective is to rely solely on the channel observations themselves. As a consequence, it is necessary to make use of unsupervised learning methods. In particular, contrastive learning has been used extensively for various tasks \cite{Jaiswal2020}. Simply put, it aims at teaching the network which samples are similar and which are different with the help of a specifically designed loss function $L$. Triplet networks (Fig. \ref{fig:triplet_network}) are an implementation of this framework where triplets of three samples each are constructed and fed to a neural network. A single triplet comprises an anchor sample $\mathbf{h}$, a close sample $\mathbf{h}^+$ and a far sample $\mathbf{h}^-$. The network produces their corresponding projections into the channel chart $\mathbf{z}$, $\mathbf{z}^+$ and $\mathbf{z}^-$ respectively. The loss function is defined as
\begin{equation}
	L(\mathbf{z},\mathbf{z}^+,\mathbf{z}^-)= \max(0,d^+-d^-+m),
\end{equation}
where $d^+=\Vert\mathbf{z}-\mathbf{z}^+\Vert_2$, $d^-=\Vert\mathbf{z}-\mathbf{z}^-\Vert_2$ and $m$ is a margin parameter. In essence, minimizing $L$ amounts to maximizing the difference between $d^+$ and $d^-$ by pulling the close sample closer and pushing the far sample farther until the difference is greater that $m$ and therefore $L=0$.

Instead of using a generic MLP as in \cite{Ferrand2021}, we propose to use the hybrid encoder presented in Section \ref{sec:hybrid_encoder} initialized with a small subset of channel observations following the proposed strategy. The parameters (i.e. weights) of the network $\boldsymbol{\theta}=\{\mathbf{D},\mathbf{Z}\}$ are then optimized through the training of the encoder in the triplet configuration. The learned charting function is denoted $\mathcal{F}_{\boldsymbol{\theta}}$.

It is important to note that constructing triplets obviously requires knowledge of positive and negative samples for a given anchor sample. In the unsupervised setting, this knowledge has to come from the dataset of inputs itself. Section \ref{sec:triplet_construction} describes how channel triplets are extracted from such a dataset. 

\section{Experiments}
\label{sec:experiments}

\subsection{DeepMIMO channels}
\begin{figure}[t!]
	\includegraphics[width=\columnwidth]{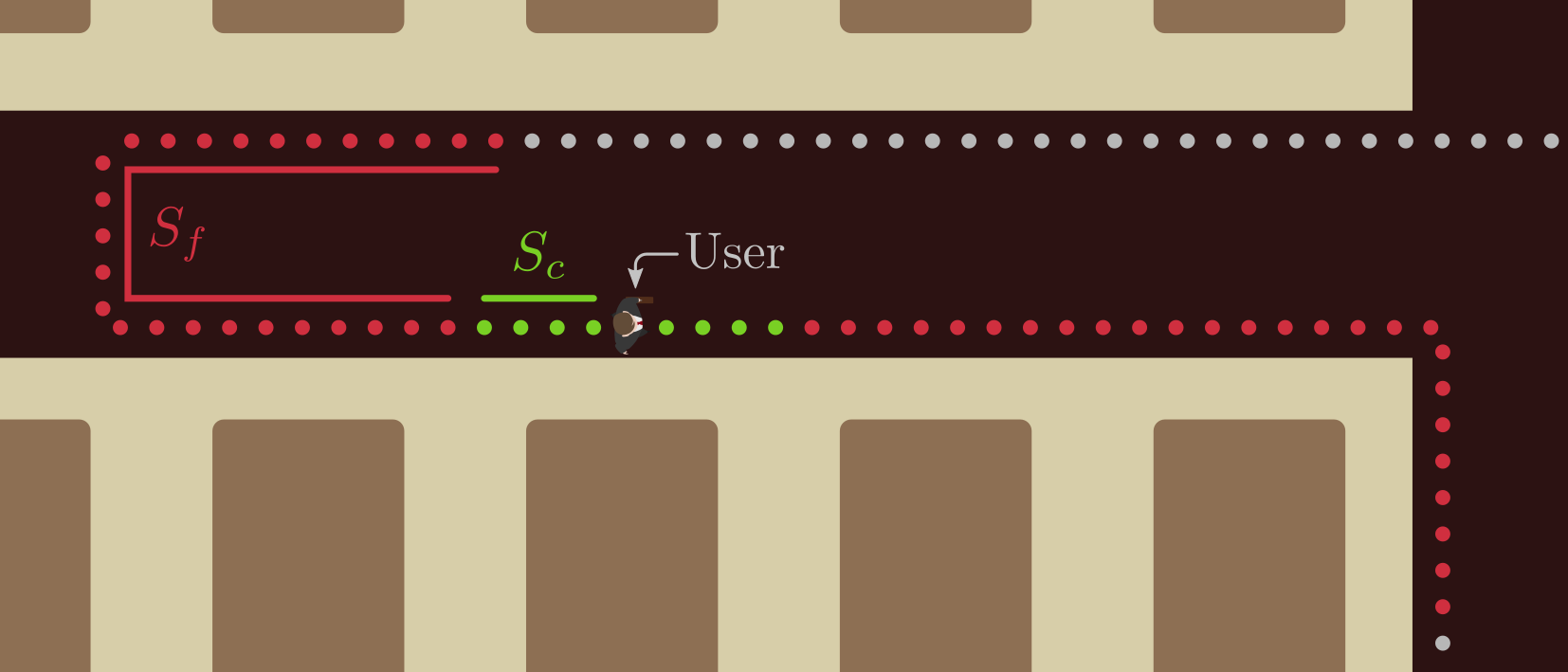}
	\caption{Locations corresponding to a triplet's close (green) and far (red) candidates. The anchor location is where the user is placed.}
	\centering
	\label{fig:dataset}
\end{figure}

To assess the performance of the proposed method, we rely on synthetic data generated from the DeepMIMO \enquote{O1} outdoor ray-tracing scenario \cite{Alkhateeb2019}. A total of $N=5910$ channels are generated and their transmit locations are geographically distributed along a path representing the route of a pedestrian (Fig. \ref{fig:dataset}). The pedestrian starts their journey at the lower rightmost point of the path and navigates counterclockwise. The pedestrian walks at a speed of $v=1.4$ m/s. The sampling rate $f_s$ is set at 7 samples/s which means that consecutive samples are separated, on average, by $1.4/7=20$ cm along the path. In addition, mild noise\comment{ (jitter?)} is added perpendicularly to the direction of travel to better emulate the walking of a real person that would not follow a perfectly straight line.\comment{ (find better wording)}

A single BS (BS 16 in the original scenario) equipped with a UPA of $N_r=64$ half-wavelength spaced antennas collects the uplink channels at a central frequency of $f_c=3.5$ GHz, with $S=16$ evenly spaced subcarriers spanning a band of 20 MHz. The resulting complex channels are of dimension $M=1024$.

\subsection{Triplet dataset construction}
\label{sec:triplet_construction}
Special care should be taken when constructing the triplet dataset as it is a critical component of the model. In the unsupervised setting, mining close and far examples w.r.t each anchor sample can be challenging in the absence of meaningful information to guide the process. However, the time-correlated nature of channel observations can be exploited. In \cite{Ferrand2021}, a timestamp-based approach is used for extracting close and far samples for each anchor sample. Indeed, each channel observation is accompanied by a timestamp corresponding to the exact moment at which it was collected by the BS. It is then legitimate to consider two observations close in time to correspond, with high probability, to transmit locations spatially close to each other. An inner temporal threshold $T_c$ can then be set to define an interval around the anchor sample's timestamp; close samples lie inside the interval and far samples outside of it. In addition, an outer temporal threshold $T_f$ can be defined to set the limits of the interval inside of which far samples lie.

The DeepMIMO dataset not providing timestamps, we rely on both the sampling rate and the temporal thresholds to deduce a threshold expressed in terms of the number of preceding and following samples w.r.t. the anchor sample. The inner and outer thresholds are expressed as 
$S_c=T_c\times f_s$ and $S_f=T_f\times f_s$ respectively. For example, with $f_s=7$ samples/s and $T_c=100 \,\text{s}$, $S_c$ amounts to 700 samples, meaning that the 700 preceding samples and the 700 following ones, distributed over a total distance of $2\times v\times T_c = 280$ m, are considered close samples for the triplet construction\comment{ (modify with final value and include the equivalent distance in meters)}. The same procedure is used to determine valid far candidates.

Knowing that samples are indexed in the order that their corresponding locations appear on the path, a single triplet $(i,j,k)$, where $i$, $j$ and $k$ are the indices of the anchor sample, the close sample and the far sample respectively, is constructed so that $i\in [0,N-1]$, $j\sim \mathcal{U}_{[i-S_c,i[\cup]i,i+S_c]}$ and $k\sim \mathcal{U}_{[i-S_f,i-S_c[\cup]i+S_c,i+S_f]}$.

Note that we can generate multiple triplets for each anchor sample simply by sampling different close and far samples, allowing us to greatly expend the dataset with no added cost.

\subsection{Performance metrics\comment{ (remove if not enough space)}}

\subsection{Results}
The evaluation of the method is done based on the contribution of three aspects: \emph{initialization}, \emph{structure} and \emph{training}. An experiment is conducted for each aspect. The base model consists of the hybrid triplet network introduced in sections \ref{sec:hybrid_encoder} and \ref{sec:triplet_loss}. The matrix $\mathbf{D}$ is initialized using a subset of $N_\text{init}=100$ channel observations randomly sampled from $\{\mathbf{h}_i\}_{i=1}^{N}$. The matrix $\mathbf{Z}$ is computed using Isomap set to keep 5 neighbors per sample. The hard thresholding nonlinearity $\text{HT}_k$ is set to pick the $k=5$ largest elements. The thresholds are set to $T_c=100$ s and $T_f=290$ s, and 5 triplets are generated for each column in $\mathbf{D}$. The margin parameter in the triplet loss is set to $m=1$. All models are trained on 70\% of the dataset for 30 epochs. Performance of the base model in particular is evaluated over 100 runs with different columns in $\mathbf{D}$ each time, as we have observed results of high variance compared to the other configurations. The evaluation is done on the 30\% unseen samples. This is in contrast to previous work where performance is evaluated on training data.

\begin{figure}[t!]
	\vspace*{0.01in}
	\includegraphics[width=\columnwidth]{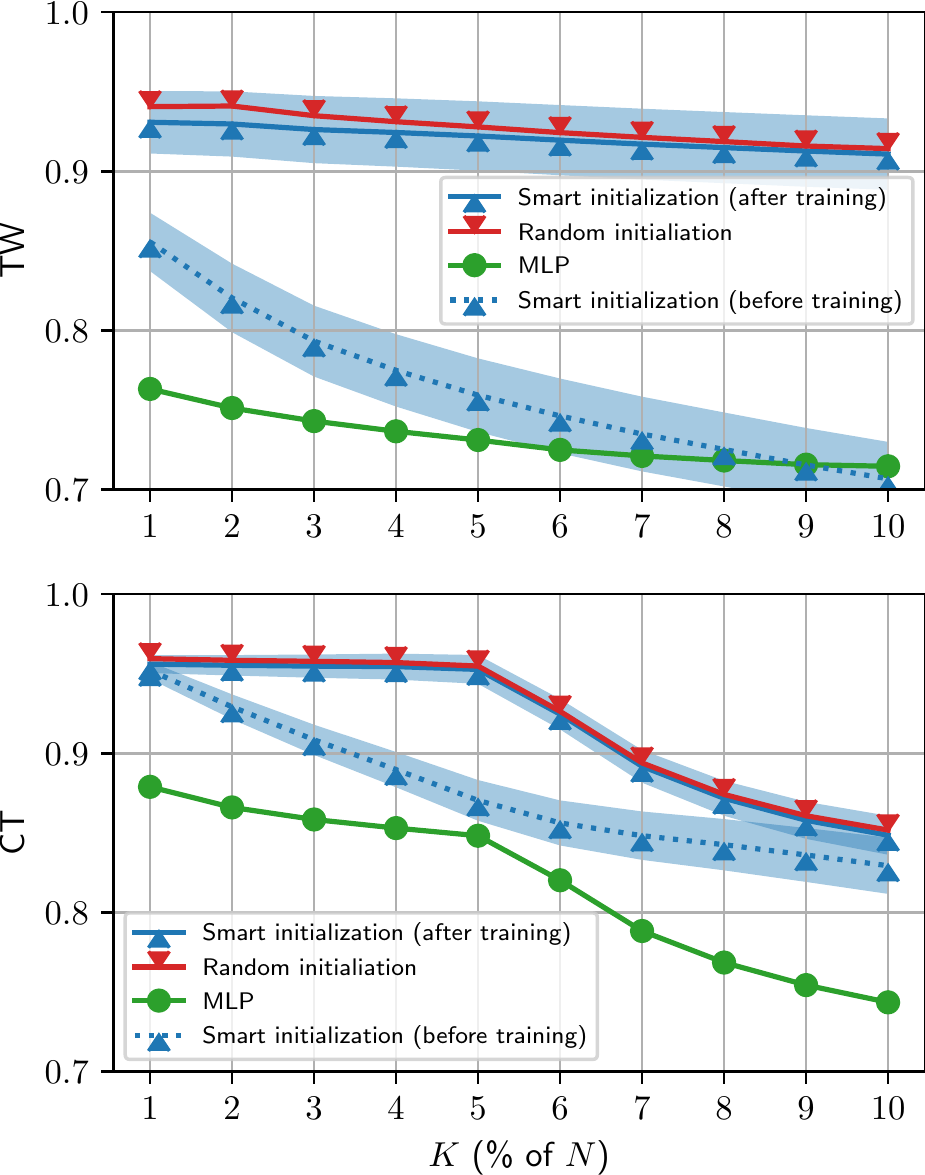}
	\caption{Comparison of the models in terms of TW and CT. Regarding the smartly initialized model, the lines represent the means over 100 runs, and the shaded regions represent the one-standard deviation bounds. Note that the lines of the smartly initialized model after training and the randomly initialized one overlap.}
	\centering
	\label{fig:lineplot}
\end{figure}

First, the base model is compared to a model with the same structure with the difference that the weight matrices are randomly initialized using the well-known Xavier initialization \cite{Glorot2010}. Fig. \ref{fig:lineplot} shows the performance of both models in terms of classical dimensionality reduction metrics, namely trustworthiness (TW) and continuity (CT) (see \cite{LeMagoarou2021} and references therein for a formal definition). We observe that the smartly initialized model, on average, achieves performance levels comparable to those of the randomly initialized one. However, results vary highly between runs, especially when it comes to TW. This means that choosing the right combination of columns to keep in $\mathbf{D}$ is critical and is something to be explored in future work. It is interesting to note that at $K=5\%$\comment{ (cutoff?)}, the CT score begins to degrade for both models. This behavior has to be linked to the thresholds $S_c$ and $S_f$ imposed during the triplet selection process.

Next, the base model is compared to an MLP composed of 6 hidden layers of sizes 1024, 512, 256, 128, 64 and 2 (output)\comment{  (check if correct)}. All layers but the last one are followed by a ReLU nonlinearity. Fig. \ref{fig:lineplot} shows that the base model performs much better both in terms of TW and CT.
Another advantage that the base model has over the MLP is the low number of parameters. The MLP has $2\,793\,600$ trainable parameters while the hybrid model only has $409\,800$ parameters. Therefore, the proposed structure is suitable for applications where computing power and time constraints are critical, which is often the case in wireless communication systems. Indeed, this model may be considered for computing channel charts on the fly to be used by subsequent tasks (e.g. handover). An additional advantage of using light models such as the one presented in this paper is the ability to train the network on the fly (or online) on the continuous stream of collected channel observations to adapt to the changing propagation environment. This aspect is yet to be explored.

Finally, the base model's performance after training is compared to its performance before training. It is clear that the training plays a significant role in improving the results for both TW and CT. Interestingly, the figure clearly shows that CT's variance in particular is reduced after training. Another important observation in favor of the smartly initialized model is that its untrained version still performs better than the trained MLP. In addition, this untrained model naturally performs better than its untrained randomly initialized equivalent (not shown on the figures, near 0.5 for both TW and CT), meaning that less training is required to achieve a desired level of performance when the model is smartly initialized.

\begin{figure*}[t!]
	\centering
	\begin{subfigure}[b]{0.23\textwidth}
		\centering
		\includegraphics[width=\textwidth]{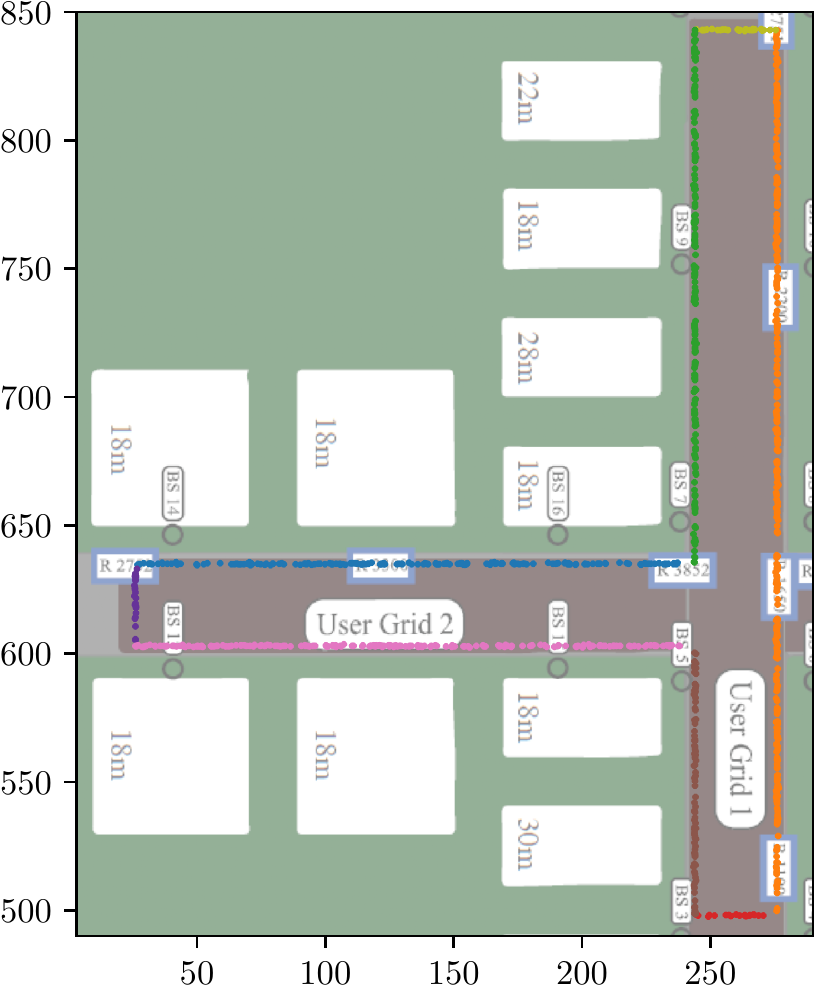}
		\caption{UE's locations}
	\end{subfigure}
	\hfill
	\begin{subfigure}[b]{0.23\textwidth}
		\centering
		\includegraphics[width=\textwidth]{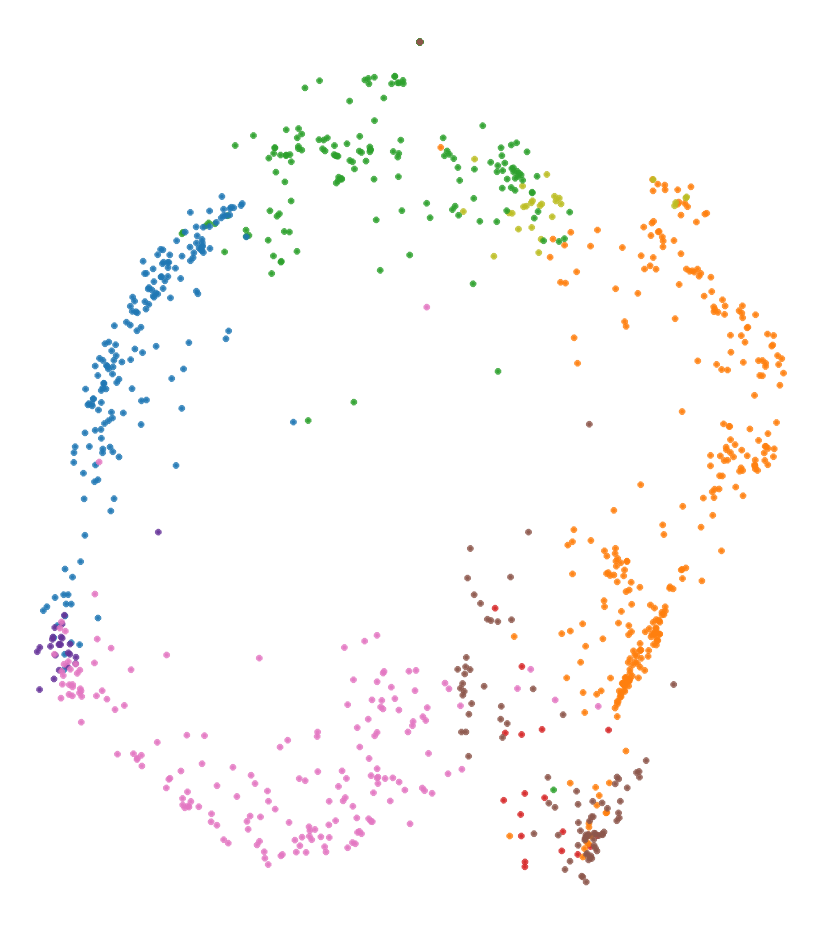}
		\vspace{-1px}
		\caption{Smart}
	\end{subfigure}
	\hfill
	\begin{subfigure}[b]{0.23\textwidth}
		\centering
		\includegraphics[width=\textwidth]{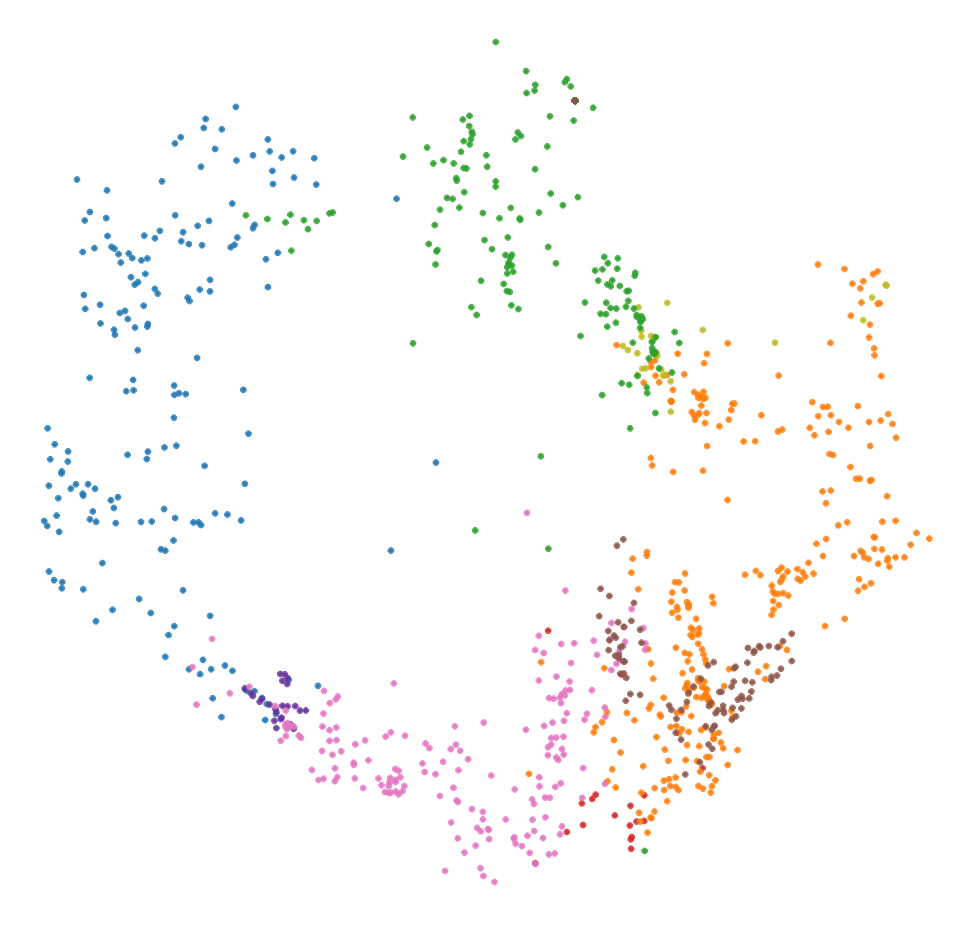}
		\vspace{8px}
		\caption{Random}
	\end{subfigure}
	\hfill
	\begin{subfigure}[b]{0.23\textwidth}
		\centering
		\includegraphics[width=\textwidth]{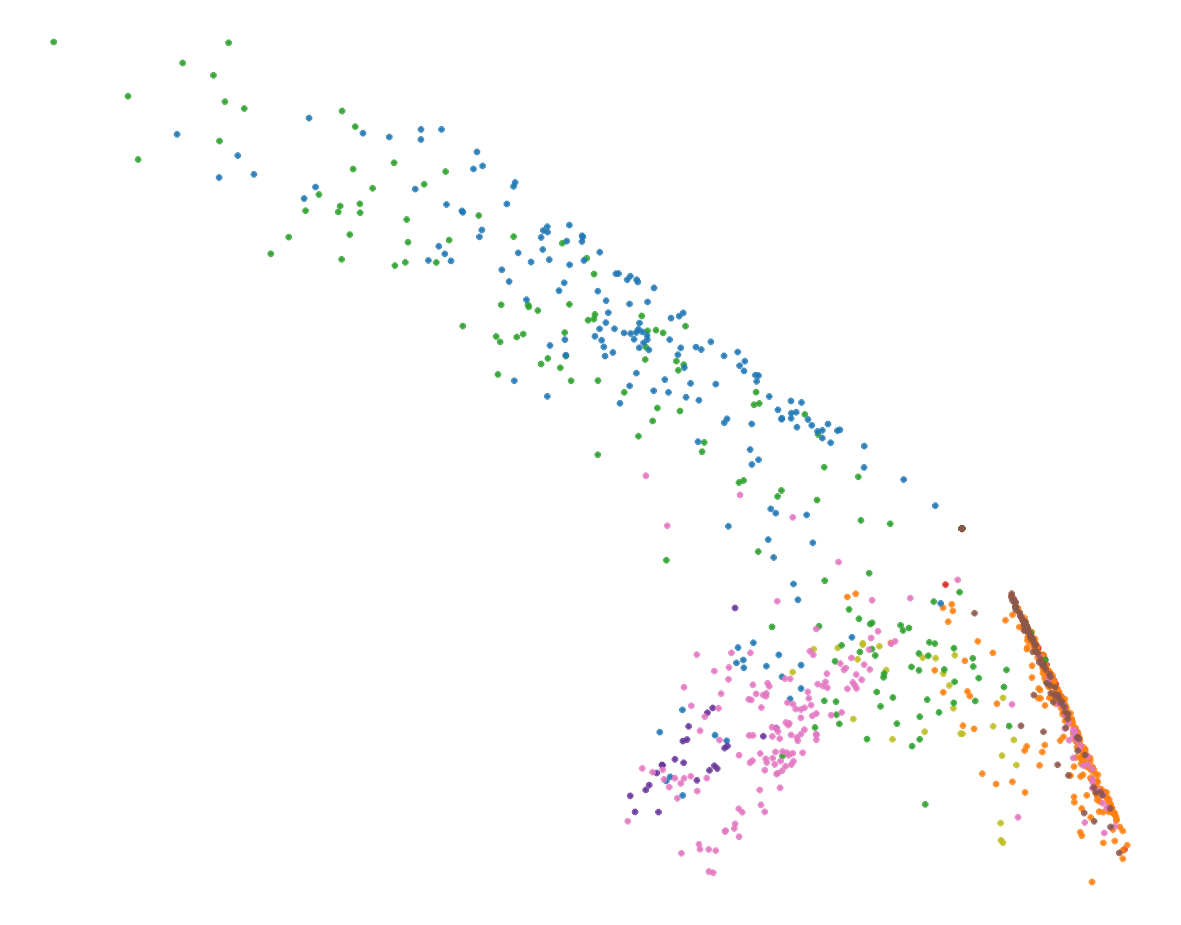}
		\vspace{6mm}
		\caption{MLP}
	\end{subfigure}
	\caption{Channel charts of the different models compared to the ground-truth locations.}
	\label{fig:chart}
\end{figure*}

Fig. \ref{fig:chart} shows the learned channel charts of the different models compared to the real physical locations on the path that the UE follows. Colors were added for better understanding of the charts. Although channel chart visualizations are subject to personal interpretation, we can nonetheless observe that local neighborhoods are more or less preserved depending on the model. For example, we observe that the chart learned by the smartly initialized model better preserves the structure of long segments of the path (blue, green, orange, brown and pink points) while points belonging to short segments are more easily mixed with the other points (purple, olive and red points).



\section{Conclusion and perspectives}
\label{sec:conclusion}
In this paper, a triplet hybrid network was proposed for the task of channel charting with the aim of reducing the computation cost, hereby allowing for on-the-fly computation and update in portable devices. The model benefits from smart initialization using a small set of channel observations, an adapted distance measure and a manifold learning algorithm like Isomap. Training the network using the triplet loss further improves performance. Furthermore, the model was shown to require fewer parameters compared to a classical MLP.

The results are encouraging and show great potential for improvement in future work. First, initialization could be improved by judiciously choosing the channel vectors to keep in the initial subset. Second, the database could be virtually augmented at no cost by constructing multiple triplets for each anchor sample by sampling different close and far samples. However, not all triplets are created equal, and some are more meaningful than others (e.g. ones containing hard-negative samples). Finding a way of selecting the optimal triplets for each sample could help with the training. Finally, extensive hyperparameter tuning could help discover\comment{  (unveil?)} a better configuration of the model.

\section*{Acknowledgement}
This work has been partly funded by the European Commission through the H2020 project Hexa-X (Grant Agreement no. 101015).

\bibliographystyle{ieeetran}
\bibliography{biblio}



\end{document}